\documentclass[sigconf]{acmart}

\usepackage[utf8]{inputenc}
\UseRawInputEncoding

\AtBeginDocument{%
  }

\copyrightyear{YYYY}
\acmYear{YYYY}
\setcopyright{acmcopyright}\acmConference[]{Conference Name}{Month date, YYYY}{}
\acmBooktitle{}
\acmPrice{15.00}
\acmDOI{xxx}
\acmISBN{xxx}

\begin{document}

\title{Experimental Investigation of Trust in Anthropomorphic Agents as Task Partners}

\author{Akihiro Maehigashi}
\affiliation{%
  \institution{Shizuoka University}
  \streetaddress{ohya 836}
  \city{Shizuoka-city}
  \state{Shizuoka}
  \country{Japan}
  \postcode{422-852}
}
\email{maehigashi.akihiro@shizuoka.ac.jp}

\author{Takahiro Tsumura}
\affiliation{%
  \institution{The Graduate University for Advanced Studies (SOKENDAI), National Institute of Informatics}
  \streetaddress{Hitotsubashi 2-1-2}
  \city{Chiyoda-ku}
  \state{Tokyo}
  \country{Japan}
  \postcode{101-8430}
}
\email{takahiro-gs@nii.ac.jp}

\author{Seiji Yamada}
\affiliation{%
  \institution{National Institute of Informatics,\\The Graduate University for Advanced Studies (SOKENDAI)}
  \streetaddress{Hitotsubashi 2-1-2}
  \city{Chiyoda-ku}
  \state{Tokyo}
  \country{Japan}
  \postcode{101-8430}
}
\email{seiji@nii.ac.jp}

\renewcommand{\shortauthors}{Maehigashi et al.}

\begin{abstract}
This study investigated whether human trust in a social robot with anthropomorphic physicality is similar to that in an AI agent or in a human in order to clarify how anthropomorphic physicality influences human trust in an agent. We conducted an online experiment using two types of cognitive tasks, calculation and emotion recognition tasks, where participants answered after referring to the answers of an AI agent, a human, or a social robot. During the experiment, the participants rated their trust levels in their partners. As a result, trust in the social robot was basically neither similar to that in the AI agent nor in the human and instead settled between them. The results showed a possibility that manipulating anthropomorphic features would help assist human users in appropriately calibrating trust in an agent.
\end{abstract}



\maketitle

\section{Introduction}
AI has been entering all aspects of life. Successful cooperation between a human user and an AI agent requires the user to appropriately adjust their use of the agent to maximize their task performance \cite{Wiegmann01, Lee04}. In the field of human factors, trust in an autonomous AI agent, such as automation, has been known to be a fundamental parameter in deciding the level of use of the agent \cite{Lee04, Parasuraman97}. 

Proper use of an AI agent is achieved through proper trust calibration, where trust in the agent is appropriately calibrated to its actual reliability \cite{Wiegmann01, Lee04}. However, people tend to over-trust and misuse the agent (inappropriate utilization of the agent) because they generally have a positive bias toward the agent, assuming that the agent performs perfectly without error. However, when people find that an AI agent has made task errors, they tend to under-trust and disuse the agent (inappropriate underutilization of the agent) because the assumption that the agent performs perfectly collapses \cite{Dzindolet02}. These poor trust calibrations would eventually lower task performance \cite{Lee04, Parasuraman97}. 

Such issues of trust calibration have been investigated and discussed in the HAI community as well \cite{Albayram20, F20, Huang17, David19}. In particular, several studies experimentally indicated that human trust in an autonomous agent and in a human differ \cite{Albayram20, Huang17}. Moreover, the over-trust and under-trust seen toward AI agents are suppressed toward humans \cite{Dzindolet02, Lewandowsky00}. These previous studies are considered to show that over-trust and under-trust toward an AI agent might be suppressed toward a social robot as in the case of humans since people tend to behave socially toward a social robot with anthropomorphic physicality that increases the sense of anthropomorphism \cite{Epley07}. 

In order to clarify how anthropomorphic physicality influences human trust in an agent, this study investigated whether human trust in a social robot with anthropomorphic physicality is similar to that in an AI agent or a human. The hypotheses in this study are as follows.\\

\noindent
H1: Trust in a social robot is similar to that in a human.\\
H2: Trust in a social robot is similar to that in an AI agent.

\section{Experiment}
\subsection{Experimental design and participants}

\begin{figure}[t]
  \centering
  \includegraphics[width=\linewidth]{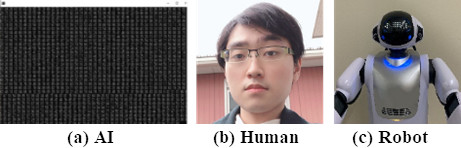}
  \caption{Task partner.}
\end{figure}

\begin{figure*}[h]
\begin{center}
  \includegraphics[width=\textwidth]{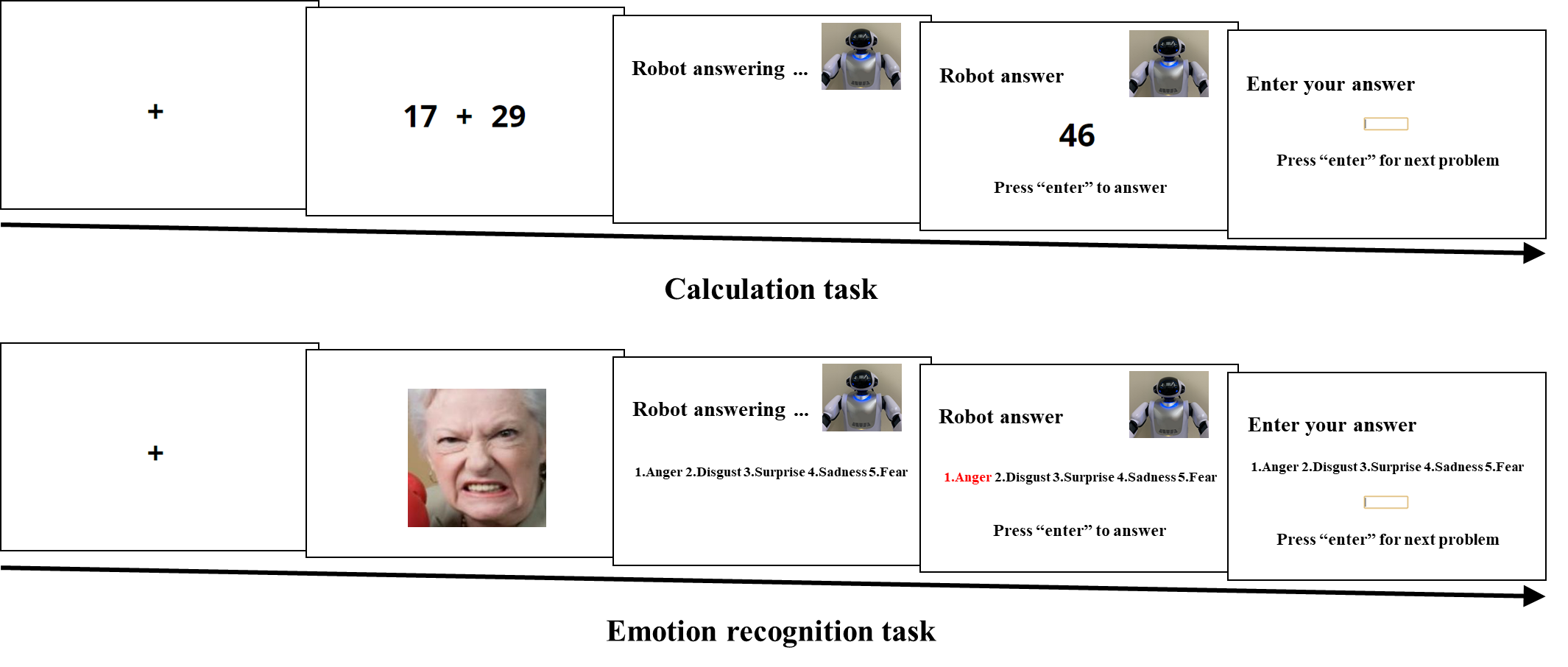}
  \caption{Procedure of calculation and emotion recognition tasks in robot condition.}
 \end{center}
\end{figure*}

The experiment had a two-factor between-participants design. The factors were the task (calculation and emotion recognition) and the partner (AI, human, and robot). A priori G*Power analysis revealed that 26 participants in each condition were needed at least for a medium effect size $(f = 0.25)$ with a power of 0.80 and alpha of 0.05 \cite{Faul07} in this experimental design. On the basis of this analysis and in consideration of the possibility that some participants would act or perform irregularly, a total of 258 participants (190 male, 68 female) were recruited through a cloud-sourcing service provided by Yahoo! Japan. Their ages ranged from 21 to 76 years old $(M = 47.24; SD = 10.49)$. They were randomly assigned to one of six conditions. As a result, in the calculation task, there were 44 participants in the AI, 42 in the human, and 45 in the robot conditions. Also, in the emotion recognition task, there were 45 participants in the AI, 41 in the human, and 41 in the robot conditions.

\subsection{Procedure}
They were randomly assigned to one of six conditions, and the task partner was introduced with one of the pictures in Figure 1 depending on the experimental condition. In the AI and robot conditions, the AI agent and social robot were explained to have computational functions in the calculation task or emotional recognition functions in the emotion recognition task and would work with the participants in real time. In the human condition, the partner was introduced as an experimental collaborator who had previously answered identical problems that would be given in the experiment.

After that, the participants first performed 10 calculation or emotion recognition problems by themselves without a partner. After that, they performed 36 calculation or emotion recognition problems with one of the partners. In the calculation task, participants mentally calculated two-digit addition problems with carry up and subtraction problems with carry down. In the emotion recognition task, participants chose which of five emotions (anger, disgust, surprise, sadness, and fear) was expressed in pictures of human facial expressions using AffectNet \cite{ali17}. 

The task procedures are shown in Figure 2. The procedure was as follows. (1) A cross was displayed at the center of the screen for 0.5 seconds, (2) a picture of a facial expression in a two-digit addition or subtraction problem in the calculation task and the emotion recognition task was presented for 5 seconds, (3) the task partner took 3 seconds to answer the problem, (4) the partner's answer was displayed, and (5) the participant's answer was entered with a numeric keypad by the participants. While the task partner was answering problems and while their answer was displayed, one of the pictures in Figure 1 was displayed depending on the experimental condition. 

Moreover, in this experiment, the partner's accuracy was manipulated to change. Each task contained 36 problems, and they were divided into 3 trials with 12 problems for each trial. The first and third trials were {\it correct trials} where the partner gave all correct answers. The second trial was an {\it error trial} where the partner gave all incorrect answers.

Regarding the measurement of trust, we asked participants to rate their trust levels during the task as in previous studies on human-automation interaction \cite{Dzindolet03, deVries03, Madhavan06}. Participants were asked ''how much do you trust your partner?'' and were required to rate their trust levels in their partners on a 7-point scale (1: Extremely untrustable - 7: Extremely trustable). The trust level was measured before the start of each task and after each of four problems.

\section{Results}
First, to confirm the analysis of the data, on the basis of the a priori G*Power analysis, we selected the data of the first 26 participants in each condition to avoid Type I and II errors in the following statistical analyses. Second, we searched for irregular data related to the accuracy rate, that is, the rate at which the participants answered correctly, without and with the partner in 2SD above or below the mean in each condition, and we eliminated the irregular data of the participants in each condition. We repeated the first and second procedure until 26 participants were secured for each condition.

\subsection{Accuracy rate}
As a task analysis, we conducted a 2 (task: calculation and emotion recognition) $\times$ 3 (partner: AI, human, and robot) $\times$ 2 (task situation: with and without partner) ANOVA on the accuracy rate in each task (Figure 3).

As a result, there was a significant interaction between the task and the task situation factors $(F(1, 150)=26.32, p<0.001, \eta_{p}^2=0.15)$. A significant simple main effect was found on the task situation factor, showing that the accuracy rate was higher with the partner than without it in the emotion recognition task $(F(1, 75)=27.80, p<0.001, \eta_{p}^2=0.27)$. There was no other significant interactions. Furthermore, there were significant main effects on the task factor $(F(1, 150)=731.49, p<0.001, \eta_{p}^2=0.83)$ and task situation factor $(F(1, 150)=25.75, p<0.001, \eta_{p}^2=0.15)$. A post-hoc G*Power analysis with an alpha of .05 revealed that the two-way ANOVA with the present sample size $(N = 156)$ obtained a power of .99 for detecting a medium effect size $(f = 0.25)$, showing satisfactory statistical power. 

This result indicated that the calculation task was easier than the emotion recognition task. The participants could show high accuracy even by themselves without a partner in the calculation task, although the participants could take advantage of performing with the partner in the emotion recognition task to increase the accuracy of their answers.

\begin{figure}[h]
  \centering
  \includegraphics[width=\linewidth]{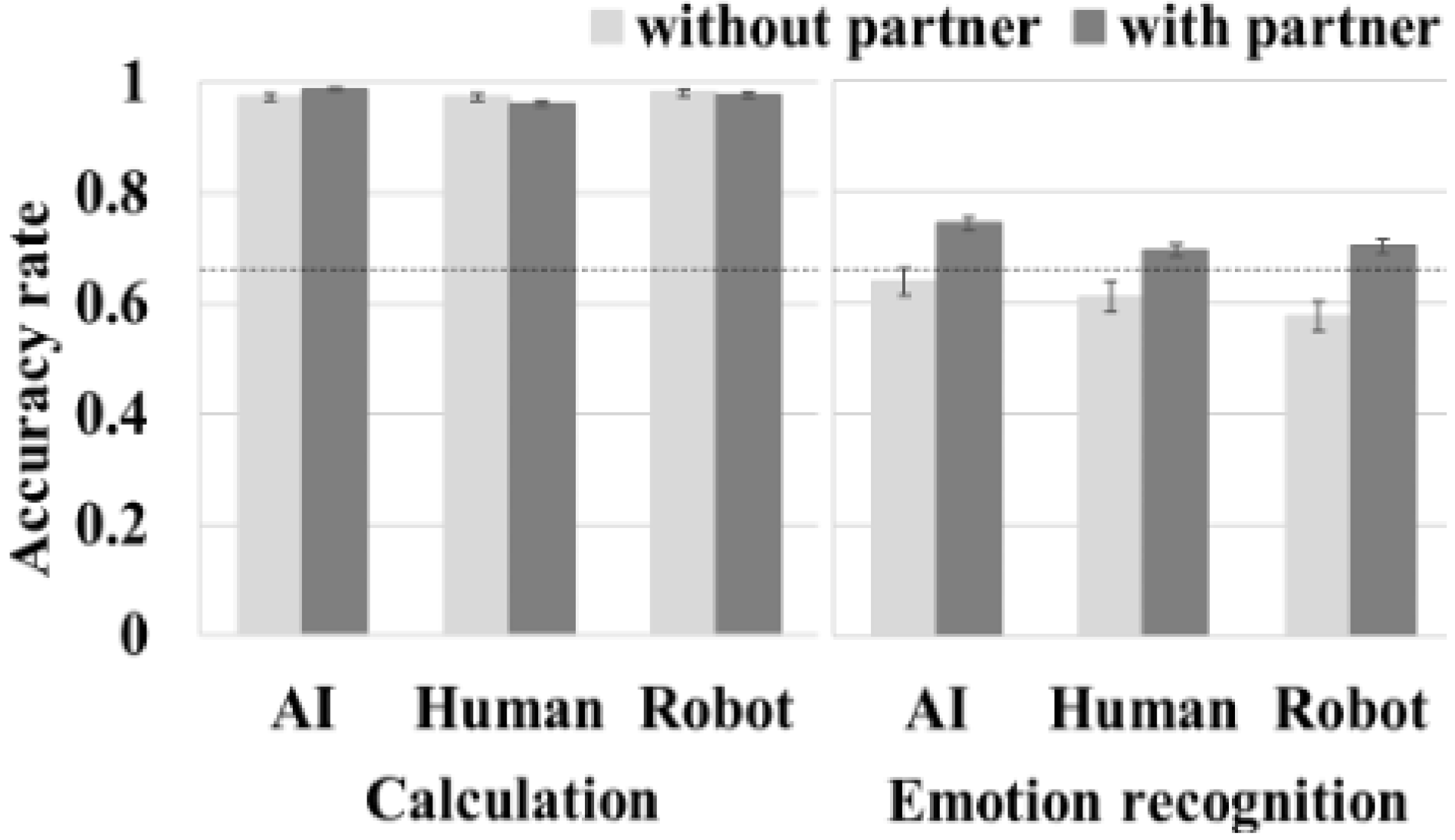}
  \caption{Accuracy rate.}
\end{figure}

\begin{figure*}[h]
\begin{center}
  \includegraphics[width=\textwidth]{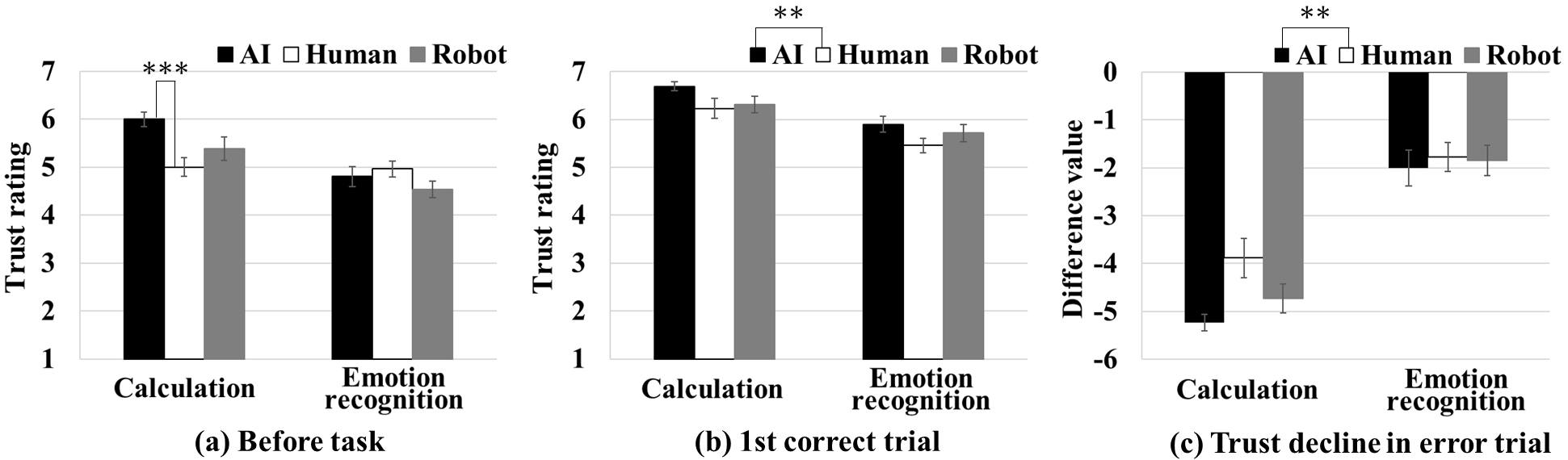}
  \caption{Trust ratings and difference values. Bars show standard errors. $** p<0.01, *** p<0.001.$}
\end{center}
\end{figure*}

\subsection{Trust rating}
As an analysis, we conducted a 2 (task: calculation and emotion recognition) $\times$ 3 (partner: AI, human, robot) ANOVA on the dependent variables in each trial. From the overall results, there were significant differences in the trust ratings among the partner conditions before the tasks and in the first correct trial (Figure 4a, b). 

First, regarding the trust rating before the tasks, there was a significant interaction ($F(2, 150)=4.21, p<0.05, \eta_{p}^2=0.05$)  and a significant simple main effect on the partner condition in the calculation task ($F(2, 150)=5.59, p<0.01, \eta_{p}^2=0.07$), showing that the trust rating in the AI condition was higher than that in the human condition ($t(150)=3.38, p<0.001, r=0.27$). 

Moreover, regarding the trust rating in the first correct trial, there was a significant main effect on the partner factor ($F(2, 150)=3.80, p<0.05, \eta_{p}^2=0.05$), and the results of multiple comparisons showed that the trust rating in the AI condition was higher than that in the human condition ($t(150)=2.73, p<0.01, r=0.22$). 

Furthermore, there was a significant difference in the decline in the trust ratings among the partner conditions in the error trial. To compare the declines in trust ratings due to partner error, we calculated the difference value between the last trust rating in the first correct trial and the first rating in the error trial for each participant and conducted an analysis on the difference value (Figure 4c). As a result, there was a significant main effect on the partner factor ($F(2, 150)=3.07, p<0.05, \eta_{p}^2=0.04$), and the results of multiple comparisons showed that the difference value for the AI condition was lower than that for the human condition ($t(150)=2.47, p<0.05, r=0.20$).

\section{Discussion and conclusion}
This study investigated whether human trust in a social robot with anthropomorphic physicality is similar to that in an AI agent or in a human in order to clarify how anthropomorphic physicality influences human trust in an agent. The results showed that the participants in this study formed trust in the social robot that was neither similar to the AI nor human and settled between them before and during the tasks. Therefore, H1 and 2 were not supported. However, the results showed that manipulating anthropomorphic features influenced trust in an agent. 

The results of this study are considered to be supportive of the human perception of agency in mind perception theory. Mind perception theory indicates that people perceive mind along with dimensions of experience (the capacity to feel and to sense) and agency (the capacity to do, to plan, and to exert self-control) \cite{Gray07, Gray12}. In regard to the agency dimension, robots with anthropomorphic physicality were perceived to have higher agency than those without it \cite{Broadbent13}; however, they were not perceived to have as much agency as humans \cite{Gray07}. Because a social robot has anthropomorphic physicality, people might perceive the robot differently from an AI agent and a human acting as a task partner and form trust in the robot differently from that in the AI agent and the human.

As a conclusion, the results showed a possibility that additional anthropomorphic features would increase anthropomorphism, and therefore, anthropomorphic physicality of an agent might be effective for suppressing over-trust and under-trust. In future work, we need to investigate how trust developed based on anthropomorphism is properly calibrated to an agent's reliability and how it could be controlled.

\begin{acks}
This work was (partially) supported by JST, CREST (JPMJCR21D4), Japan.
\end{acks}

\bibliographystyle{ACM-Reference-Format}
\bibliography{hai22-43}

\appendix

\end{document}